# Space-based Measurements of Neutron Lifetime: Approaches to Resolving the Neutron Lifetime Anomaly

David J. Lawrence, Jack T. Wilson, and Patrick N. Peplowski

Johns Hopkins University Applied Physics Laboratory
Laurel, MD 20723

**Abstract**

Free neutrons have a measured lifetime of 880 s, but disagreement between existing laboratory measurements of ~10 s have persisted over many years. This uncertainty has implications for multiple physics disciplines, including standard-model particle physics and Big-Bang nucleosynthesis. Space-based neutron lifetime measurements have been shown to be feasible using existing data taken at Venus and the Moon, although the uncertainties for these measurements of tens of seconds prevent addressing the current lifetime discrepancy. We investigate the implementation of a dedicated space-based experiment that could provide a competitive and independent lifetime measurement. We considered a variety of scenarios, including measurements made from orbit about the Earth, Moon, and Venus, as well as on the surface of the Moon. For a standard-sized neutron detector, a measurement with three-second statistical precision can be obtained from Venus orbit in less than a day; a one-second statistical precision can be obtained from Venus orbit in less than a week. Similarly precise measurements in Earth orbit and on the lunar surface can be acquired in less than 40 days (three-second precision) and ~300 days (one-second precision). Systematic uncertainties that affect a space-based neutron lifetime measurement are investigated, and the feasibility of developing such an experiment is discussed.

## 1. Introduction

Free neutrons are unstable and have a mean lifetime, $\tau_n$, of ~15 minutes for beta decay (via the weak interaction) into a proton, electron, and antineutrino. Knowledge of the neutron lifetime is important because its precise value provides constraints to both fundamental physics and Big-Bang nucleosynthesis. There currently exist two types of laboratory experiments – "beam" and "bottle" – that have measured neutron lifetime. Beam experiments involve measuring the protons or electrons resulting from beta decay of cold neutron beams, and bottle experiments use storage (material, magnetic and/or gravitational) to trap neutrons and measure the number surviving as a function of time. While the reported results from both types of experiments have relative errors of less than 1%, there exists disagreement of up to five standard deviations ($\sigma$) between individual results and the average [1], and the averages of the beam (888±2 s) and bottle (879.5±0.5 s) measurements disagree by 8.5 s. The disagreement between these two types of experiments may be the result of systematic uncertainties that are not fully understood and/or may indicate the existence of some physical processes that are not yet explained.

In light of this discrepancy, a third technique of space-based neutron spectroscopy was put forward by Feldman et al. [2] to measure neutron lifetime. This technique measures neutrons from planetary surfaces or atmospheres that are generated by nuclear spallation reactions when galactic cosmic rays (GCRs) impact the planets. The primary spallation neutrons have initial energies, $E_n$, greater than 1 MeV, but a fraction of them are down scattered in energy via inelastic scattering collisions with atmospheric and/or surface material prior to escaping into space. The resulting energy spectrum for neutrons that escape into space consists of a power law shape for energies in the range of a few eV to ~500 keV (epithermal neutrons), and a Maxwellian shape for the lowest energy (thermal) neutrons that are in thermal equilibrium with the surrounding material (e.g., [3] and Figure 1). For large enough planetary bodies (e.g., like the Moon or bigger), thermal neutrons are gravitationally bound and have a surface-to-surface time-of-flight that is comparable to the neutron lifetime. Large asteroids have masses that are two orders of magnitude smaller than the Moon, and thus cannot gravitationally trap even the slowest neutrons (see Section 2.1).

The first measurement of neutron lifetime using this space-based technique was reported by Wilson et al., [4] using neutron data collected by NASA's MESSENGER mission during flybys of Venus and Mercury in 2007 and 2008. While the lifetime measurement using these data of 780±60 $_{stat}$ ± 70 $_{syst}$ s does not have an uncertainty low enough to be competitive with existing laboratory measurements, it nevertheless demonstrated the feasibility of the space-based technique. Another study using data from the Lunar Prospector mission at the Moon has reported an independent measurement of neutron lifetime $900^{+40}_{-50}{}_{\text{stat}} \pm 17_{\text{syst}}$ s [5]. Additional space-based data from Mars, Mercury and the Moon [6, 7] exist with which further measurements of $\tau_n$ may be attempted. However, none of these existing datasets were generated with instruments designed for the purpose of a neutron lifetime measurement. These experiments therefore have constraints that limit their usefulness for neutron lifetime measurements, either because of limited statistical precision and/or large systematic uncertainties. With the knowledge gained by planetary neutron measurements in recent years, we have better information for how a space-based neutron lifetime experiment could be realistically designed and built within the constraints of a small-to-moderate space mission.

A number of factors need to be considered when planning such a mission. Measurement constraints such as the host planet and measurement configuration (e.g., circular orbit, elliptical orbit, landed) have a primary importance for how to carry out a measurement with the optimum statistical precision, as well as minimizing and mitigating systematic uncertainties. Implementation factors, such as instrument heritage and the ability to reach a given location, affect the practicality of accomplishing a mission scenario. Finally, detector considerations such as type ($^3$He gas, lithium glass, borated plastic, etc.), geometry, and number of detectors also play an important role. Within these constraints we investigate two types of measurements: planetary orbital and planetary landed. For orbital missions, we consider the Earth, the Moon, and Venus. Each of these bodies are large enough to have sufficient gravity to make neutron lifetime measurements, and all are reasonably accessible with standard space launch capabilities. For landed measurements, we consider the possibility of neutron lifetime measurements made from a single location on the lunar surface. A primary reason for investigating measurements from the lunar surface is that a renewed effort to place low-cost instrumentation on the Moon might enable a neutron lifetime experiment to be feasibly emplaced there.

The organization of this paper is the following. Section 2 discusses orbital neutron measurements, provides a baseline instrument design, and describes the expected performance of orbital neutron lifetime measurements at the Earth, the Moon, and Venus. Section 3 describes how a neutron lifetime measurement could be made from the lunar surface. Section 4 gives a general discussion of the systematic uncertainties associated with all space-based neutron measurements.

Section 5 gives a summary of these results and discusses the next steps needed for fielding a realistic neutron lifetime mission.

## 2. Orbital Measurements

### 2.1 Principal of Technique

Planetary neutron measurements have been made at seven different planetary bodies, which include: the Moon, Mars, Mercury, Venus, asteroids 4 Vesta and 1 Ceres, and the Earth. For the non-Earth measurements, neutrons have been used to quantify surface or atmospheric composition [e.g., 7, 8, 9-11]. Based on these measurements and the modeling carried out to support these measurements [e.g., 7], there is a good understanding of the effects of planetary neutron generation, transport, and detection by orbital instruments, as well as the neutron flux dependence on surface and/or atmospheric composition. Figure 1 shows modeled neutron energy flux spectra for the Moon, Earth, and Venus. These spectra were generated using the particle transport code MCNPX, which simulates GCR-induced neutron production and transport, including effects from surface and/or atmospheric composition, surface temperature, and gravitational binding. For the Moon, two spectra are shown – one for lunar highlands materials that have low abundances of thermal-neutron absorbing materials Fe, Ti, Sm, and Gd; and one for lunar maria materials that have high abundances of these elements. The effects of these abundance differences are seen primarily for thermal energies ($E_n < 0.4$ eV), where the maria materials have substantially fewer thermal neutrons than highland materials due to neutron absorption. These and other compositional effects have been quantitatively validated using orbital measurements [e.g., 3, 12]. The neutron spectra for Earth and Venus were generated using a 50-layer atmosphere model that accounted for density and temperature in each layer, as well as standard compositional values [13]. The neutron flux for the Earth is significantly lower than either the Moon or Venus because Earth's atmosphere is dominantly composed of nitrogen (78%), which is a relatively strong neutron absorber. In contrast, Venus' atmosphere is dominated by carbon and oxygen (96.5% by volume $CO_2$), and both those elements have a relatively small thermal-neutron-absorption cross section. As a consequence, at the peak energy of 0.04 eV, the thermal neutron flux at Venus is almost a factor of 150 larger than that of Earth.

Neutron lifetime measurements are performed via measurements of thermal neutrons, which follow trajectories that are significantly influenced by the planet's gravitational field, with sufficiently low-energy neutrons being gravitationally bound to the planet. Feldman et al. [14] derived the time required, Δt, for a neutron to leave and then return to a planetary surface:

$$\Delta t = \frac{R_M (m/2V)^{1/2}}{[1-(K/V)]^{3/2}} \left[ \left(\frac{\pi}{2}\right) + A + \sin^{-1}\left(\frac{B}{[A^2+B^2]^{1/2}}\right) \right] \quad (1)$$

where,

$$A = \left[ 4\left(\frac{K}{V}\right)\left(1-\frac{K}{V}\right)\mu^2 \right]^{1/2} \text{ and } B = (2K/V - 1), \text{ for } K/V < 1 \quad (2).$$

In this expression, the neutron kinetic energy is $K$ (in eV), the planetary radius is $R_M$, neutron mass is $m$, and the neutron launch direction is $\mu = \cos(\theta)$, where $\theta$ is the angle relative to the surface normal. The gravitational binding energy is $V = GMm/R_M$, where $G$ is the gravitational constant, and $M$ is the planetary mass. The values of $V$ for the Moon, Venus, and Earth are 0.029 eV, 0.560 eV, and 0.653 eV, respectively.

Figure 2 shows Δt versus energy for the Moon, Venus, and Earth using two different launch directions of $\theta = 0°$ (solid lines) and $\theta = 60°$ (dashed lines). The return times have values comparable to the mean neutron lifetime, which, in part, enables the space-based measurement of neutron lifetime. Figure 2 illustrates key differences between the planetary bodies in regards to thermal neutrons. Since Earth and Venus have similar masses (Earth has only a 20% larger mass than Venus), neutron return times are also similar, and gravitationally bound neutrons include all thermal energies (compare with Figure 1b). In contrast, the mass of the Moon is lower than Venus and Earth by factors of 66 and 81. Therefore, lower energy neutrons can escape the Moon's gravity and only a fraction of thermal neutrons are gravitationally bound. In comparison to the Moon, the largest asteroids –1 Ceres and 4 Vesta – have neutron gravitational binding energies of 3.4 x $10^{-4}$ eV and 1 x $10^{-5}$ eV, respectively; thus, thermal neutrons are not gravitationally bound to these and smaller bodies.

*2.2    Measurement Implementation*

There are multiple ways to make space-based neutron measurements. These include using $^3$He gas proportional counters [15] and different types of scintillators such as borated plastic (BP) and Li glass (LG) [16]. In addition to different sensors, there are different methods for separating thermal from higher energy neutrons. A straightforward energy discrimination is to cover neutron sensors with thermal-neutron absorbing materials (e.g., Cd) that enable count-rate differences between thermal and non-thermal-neutron detecting sensors to be a measure of thermal neutrons [15]. Another method for discriminating neutron energies with orbital neutron measurements is the Doppler filter technique [17]. This technique takes advantage of the similarity in velocity between typical spacecraft in orbit about a planet (few km/s) and thermal neutrons (a thermal neutron with energy of 0.025 eV has a velocity of 2.2 km/s). An enhancement of thermal neutrons is measured when the spacecraft velocity vector is parallel with the sensor normal vector; a relative decrease in thermal neutrons occurs when the spacecraft velocity vector is anti-parallel to the sensor normal.

We posit a baseline neutron lifetime instrument design based on experience with the MESSENGER Neutron Spectrometer (NS). Figure 3 illustrates the MESSENGER NS, which measured a range of neutron energies using a BP scintillator and two LG scintillators. The BP sensor was a 10 cm$^3$ cube of plastic scintillator. The two LG sensors were 10 cm by 10 cm by 4 mm LG plates placed on two opposite sides of the BP sensor. Each sensor was read out by separate photomultiplier tubes (PMTs). Thermal and epithermal neutrons were measured with the two LG sensors using the Doppler filter technique [18].

Figure 4 illustrates how the energy discrimination is accomplished with the Doppler filter technique by showing a particular orbit of the MESSENGER spacecraft around Mercury. In the normal MESSENGER operations, the spacecraft was in an eccentric orbit around Mercury with a periapsis altitude from Mercury of a few hundred km, and an apoapsis altitude of ~10,000 km [19]. Note that for the MESSENGER mission, each 12- or 8-hour orbit had a large variety of pointing attitudes in order to satisfy the various observational requirements of the mission's seven different instruments [19]. The neutron count rate from the LG2 sensor is shown in Figure 4c, where the increasing count rate is due to the larger neutron flux at close distances to Mercury's surface. Due to the fact that the neutron sensors are largely hemispherical detectors, measured neutron counts are statistically significant when the detector is within one-body radius (2438 km) of Mercury's surface. The altitude and velocity-direction dot product with the spacecraft x-axis direction are shown in Figure 4a and 4b, respectively.

For the orbit shown in Figure 4, the spacecraft executed a rotation near the periapsis such that the velocity direction went from being aligned with the normal to the LG2 sensor to being anti-aligned with the LG2 sensor. This rotation is illustrated in Figure 4b with the vector-velocity dot product with the spacecraft x-axis (which is aligned with the LG2 normal vector). The effect of the rotation is seen where there is a relatively large thermal-neutron count rate prior to the rotation, and a count-rate drop after the rotation. Note that the spacecraft velocity of 3.5 km/s at periapsis corresponds to a neutron energy of 0.065 eV.

The effects varying neutron lifetimes would have on the neutron count rate is shown by simulated count rates (colored traces) in Figure 4c. The details of how these simulated count rates were implemented is described in various prior studies [4, 7, 11, 18, 20, 21], with the foundational algorithms given by [14]. Neutron lifetime effects manifest as count-rate variations for different $\tau_n$ values when thermal neutrons are detected prior to the rotation. Specifically, short lifetimes yield lower count rates and long lifetimes yield higher count rates. In contrast, post-rotation lifetime-derived variations are suppressed since the LG2 sensor only measures higher-energy epithermal neutrons that are not noticeably affected by lifetime. The relative count-rate difference between pre- and post-rotation measurements therefore provides a measure of $\tau_n$. We note that while this particular orbit provides an optimum configuration for a $\tau_n$ measurement (and well illustrates the technique), many similar such orbits are needed to achieve a statistically significant measurement.

Based on this one-orbit scenario, Figure 5 shows a notional detector arrangement that could carry out a sequence of orbits to achieve a statistically significant neutron lifetime measurement. This design uses four 100 cm$^2$ by 4 mm thick LG detectors arranged around an axis of rotation that is perpendicular to the spacecraft velocity direction around a planet. For an appropriate rotation speed (e.g., one rotation per few minutes), each detector will cycle through measurements ranging from thermal to epithermal energies. The four detectors provide an increase in statistics over a single detector, as well as redundancy if a detector fails. The four identical detectors in a constant rotation also allows a consistent detector-to-detector normalization to account for time dependent count-rate changes, such as from time variable cosmic ray flux variations (see Section 4).

In regards to engineering considerations, this baseline design has a number of benefits. Most important, the sensor technology has very high spaceflight heritage, as the identical scintillator and photomultiplier tubes used for MESSENGER could be used here. The four-sensor configuration is ideally suited for existing four-channel electronics systems being used on planetary neutron gamma-ray and neutron experiments planned for launch in the near future [22, 23]. Finally, this sensor arrangement is appropriate for implementation on a small satellite, which can reduce neutron background and systematic measurement effects (see Section 4), as well as lower satellite design and launch costs.

*2.3    Neutron Lifetime Measurements at the Earth, Moon, and Venus: Statistical Uncertainty and Required Measurement Time*

Here we explore mission scenarios using the detector arrangement in Figure 5 to estimate the statistical performance for possible measurements at the Earth, Moon, and Venus. Earth orbit is the most easily accessible location for a space-based neutron detector, and was the location assumed by Feldman et al. [2] in the initial study that proposed space-based neutron lifetime measurement. While not as accessible as low-Earth orbit, lunar orbit is visited by many spacecraft, and with renewed interest in lunar exploration will likely be a more frequent destination by spacecraft. Finally, while Venus orbit is more difficult to reach than Earth or lunar orbit, it is nevertheless visited with some frequency either for dedicated missions or gravity assist flybys. For any of these locations, a neutron lifetime mission could be flown as a stand-alone mission or as a hosted payload for another mission.

To understand the fundamental statistical performance of a detector in orbit about these planetary bodies, we first simulated a circular orbit about each planet. For Earth and Venus, we assume an altitude of 500 km; for the Moon due to its smaller size, we assume an altitude of 50 km. These are straightforward altitudes to obtain for orbiting spacecraft around these bodies. We also simulated elliptical orbits at Venus with periapsis and apoapsis altitudes of 250 km and 750 km, respectively. This orbit keeps approximately the same average 500-km altitude as the circular orbits. For this simulation, we assumed the four sensors are rotating around the sensor center line with a rate of 0.5 rotations per minute. Figure 6a shows the simulated count rates for one orbit using one of the four sensors with different assumed $\tau_n$ values. The longer time scale (zero to 50-minute) count-rate variation is due to the altitude-dependent count rate. The higher time frequency count-rate changes are due to the sensor rotation that constantly varies from Doppler-enhanced, with a larger fraction of thermal neutrons, to Doppler-suppressed, with a larger fraction of epithermal neutrons. Figure 6b focuses on 20 minutes of the orbit where the count rates from two of the opposing sensors are shown.

To estimate the statistical precision for each planet and orbit type, a $\chi^2$ value was calculated comparing each of the models to a 900-s reference model, which was used as a proxy for the measurements, and assuming Poisson statistics on the 'measurement'. Given the $\chi^2$ estimate for one orbital period $T$, the expected $\chi^2$ value after a total measurement time $t$ was calculated by scaling with a proportionality factor $\chi^2(t) = t/T \, \chi^2(T)$. Thus, the expected value of $\tau_n$ that is distinguishable from the reference lifetime $\tau_n^{900}$ after a total measurement time $t$ is given by $\chi^2(t, \tau_n) - \chi^2(t, \tau_n^{900}) = 6.63$, where the 6.63 is the 3-sigma significance for a $\chi^2$ difference. The expected uncertainty is then $\sigma(\tau_n) = \tau_n - \tau_n^{900}$. In practice, a discrete set of different lifetimes were considered in the models (900, 901, 905, 910, 950, 1000, 1500 s) with the lifetime determined by logarithmic interpolation between these models. The results of these calculations are shown in Figure 7.

An uncertainty of $\sigma_\tau = 3$ s is required (and 1 s desired) to provide discrimination between the two laboratory-based techniques. At Venus, this level of statistical uncertainty is achieved in less than 1 day for $\sigma_\tau = 3$ s, and less than 4 days for $\sigma_\tau = 1$ s. In Earth orbit, these levels of precision require less than 40 days and ~300 days, respectively; in lunar orbit, precisions of 3 s and 1 s require more than 100 days and ~1000 days (2.7 years), respectively. Equivalently, 20 days of observation time would result in a measurement uncertainty of $\sigma_\tau = <1$ s at Venus, $\sigma_\tau = 4$ s at the Earth, and $\sigma_\tau > 7$ s in lunar orbit. The data points in Figure 7 show the measured uncertainty for the two-existing space-based neutron lifetime measurements at Venus and the Moon [4, 5]. Both measured values are above the curves as they should be, with the MESSENGER measurement considerably above. The MESSENGER measurement is less constraining than suggested by the results in Figure 7 as only two detectors were present and the mean altitude during the flyby was several thousand km instead of 500 km, with a corresponding smaller count rate. The Moon measurement is not directly comparable because it used different detectors ($^3$He sensors) than modeled here, yet the detector areas for the two measurements are similar at around 100 cm$^2$.

There are two primary reasons for the large difference in required observation times among Venus, Earth, and the Moon. Most importantly, Venus' atmospheric composition with a large $CO_2$ abundance and low $N_2$ abundance results in a very high flux of thermal neutrons compared to the Moon and especially the Earth (Figure 2). Second, because both Venus and the Earth have a larger gravitational field than the Moon, a larger fraction of thermal neutrons are gravitationally bound, thus enabling a more robust $\tau_n$ measurement. Thus, based on these results, we conclude Venus is the optimum location for carrying out orbital $\tau_n$ measurements. However, within a one-year mission duration, statistically significant measurements could also be carried out in Earth orbit. Further, if a larger detector could be flown at any body, the required accumulation time shown in Figure 7 could be reduced by the fractional area increase of detector size.

## 3. Surface Based Neutron Lifetime Measurements on the Moon

There is a recent effort with international space agencies to return people as well as scientific and engineering instrumentation to the lunar surface. With the recognition that this "return to the Moon" strategy provides an opportunity to place a neutron lifetime experiment on the lunar surface, here we investigate a notional experiment and how well it could measure $\tau_n$.

In keeping with the sensor architecture of the orbital experiment, here we suggest a four-sensor experiment that enables the separate measurement of upward and downward thermal neutrons (Figure 8). To maximize detection area, we use flat Li-glass sensors (10 cm by 10 cm by 4 mm), where a pair of the sensors look upwards and a pair look downwards. One of each pair is covered in Cd to block out thermal neutrons; the other of the pair is bare, thus allowing detection of both thermal and epithermal neutrons. The difference between the measured count rates of both detectors is the thermal-neutron-only rate. Each pair of sensors is then placed within shielding material – here we assume $^{10}$B-enriched boron carbide ($B_4C$) – to reduce background detections of wide-angle thermal and epithermal neutrons. While the exact size and thickness of shielding material would eventually be traded against experiment mass and performance, here we assume a thickness of 5 cm. One pair of sensors is configured to observe the upward flux of neutrons, and the other pair to observe the downward flux. The optimum location for placing such an experiment would be where there are low abundances of thermal-neutron-absorbing elements such as Fe, Ti, Gd, and Sm, thus providing the highest fluxes of thermal neutrons on the Moon [3]. Figure 9 shows the respective upward and downward neutron fluxes for an experiment placed on a location with low abundances of neutron absorbing materials. Since the Moon's gravity only traps a fraction of the thermal neutrons escaping from the lunar surface, this downward flux is relatively small compared to the upward flux.

With this sensor arrangement, we have determined the statistical precision that can be achieved for measuring $\tau_n$ as a function of accumulation time using the same procedure that was described in Section 2.3 for orbital measurements. The black dashed line in Figure 7 shows the results, where a landed lunar measurement is statistically similar to an Earth orbiting measurement. The primary reasons for the improvement in statistical precision over that of a Moon orbiting measurement are the increase in count rate due to being closer to the planet and the clean isolation of low-energy, gravitationally bound, neutrons. For one lunar sunlight period (half a lunar day, or 15 Earth days), a measurement precision of ~4 s could be achieved, with slightly better than 3 s in a full lunar day. Longer-term (multi-month) operation on the lunar surface will require accommodations to survive the cold lunar night. However, if such a detector were accommodated with multi-month operation, then a 1-s-precision measurement could be acquired in less than a year of accumulation time. We therefore conclude that a landed lunar measurement is feasible within existing technologies.

## 4. Expected Systematic Uncertainties

In Section 3, we showed that it is possible to make high-precision measurements of $\tau_n$ with both orbital and landed lunar measurements. However, the driving factor for achieving a competitive $\tau_n$ measurement is likely not statistical precision, but systematic uncertainties. Ultimately, the measurement of $\tau_n$ will be obtained by generating models of the expected count rates for given $\tau_n$ values, and then comparing these simulations with the measured data. The degree to which this model/data comparison can accurately measure $\tau_n$ depends on how well the simulations account for all the non-$\tau_n$-dependent factors. Uncertainties in these factors are the systematic uncertainties that need to be understood in order to acquire a competitive $\tau_n$ measurement.

Table 1 lists six different classes of systematic uncertainties that can affect the $\tau_n$ measurement, identified in part via the previous neutron lifetime analysis efforts using MESSENGER and LP data. Fully understanding the extent and magnitude of these uncertainties is beyond the scope of this paper. A detailed study and accounting for these uncertainties is likely needed for most, if not all, of the different classes of systematic uncertainties. Here, we outline each type of uncertainty, and describe the mitigations that can be employed, as well as additional work needed to ensure these are effective mitigations.

*Galactic cosmic ray (GCR) flux:* GCRs initiate the interactions that generate neutrons, from which $\tau_n$ is measured. The GCR flux can vary in time by a significant amount (few to tens of percent) over time scales of hours to days to months [24, 25]. Without accounting for these variations, they could easily mask any neutron lifetime measurement. There are multiple ways the GCR time variation can be taken into account. Separate GCR monitors could be included on an orbital or landed payload to make an independent measurement of GCR time variability. Alternatively, the sensors making the neutron measurements can also monitor the GCR time variation. For the orbital package shown in Figure 5, GCR time variation can be monitored using the non-Doppler-enhanced epithermal neutron measurement. This is done by virtue of the fact that epithermal neutrons will respond to time variations in GCRs, but are not sensitive to $\tau_n$. Similarly, for the landed measurement scenario (Figure 8), GCR time variations can be monitored by virtue of one set of sensors measuring epithermal neutrons, and the other measuring thermal plus epithermal neutrons. We also note that for Earth orbiting measurements, atmospheric neutron production from the GCR varies as a function of latitude due to Earth's magnetic field, which cuts off lower-energy GCRs near the equator and allows lower-energy GCRs near the poles. Thus, GCR monitoring in an Earth-based measurement needs to account for this variability. A dedicated fast-neutron monitor is one way to measure such a latitude-dependent parameter [2].

A number of tasks need to be carried out to ensure GCR time variability introduces a small systematic uncertainty to a $\tau_n$ measurement. These include deriving the statistical uncertainty of an expected GCR count rate from an independent GCR monitor, if one is chosen for use on the mission. Alternatively, the statistical precision of epithermal neutron derived GCR rates, and the corresponding derived $\tau_n$ uncertainty can be determined in the different orbital and landed measurement scenarios. Such information can be determined using prior measurements of GCR variability with similar sensors [25].

*Atmospheric/surface composition:* Knowledge of the elemental composition of the gaseous atmosphere or solid surface is needed to derive a $\tau_n$ measurement. The reason is that different elemental compositions can cause variations in thermal neutron fluxes. At the Moon, the measured thermal neutron flux varies by over a factor of three over the entire lunar surface [26].

Gaseous atmospheres will likely have less spatial variability in thermal neutron flux due to their homogeneous nature; however, the elemental composition still needs to be known to derive $\tau_n$. Venus provides an ideal atmosphere as it has only two primary constituents of $CO_2$ and $N_2$, with trace amounts of other species. The impact that these trace elements may have on systematic uncertainties at the 1-to-3 s precision level needs further investigation. In contrast, the elemental composition of Earth's atmosphere is better known than Venus' atmosphere. But the Earth contains more dominant constituents (e.g., variable $H_2O$ that can affect the neutron flux), which may prevent knowing the atmospheric composition with sufficient precision to make a competitive $\tau_n$ measurement.

For a landed lunar measurement, the elemental composition of the landing site can be determined using existing lunar measurements [e.g., 27]. In addition, it is possible that a small gamma-ray sensor could be included on a landed sensor package to measure the abundances of the elements that are most important to characterize the near-sensor neutron absorption. Due to the long distances traveled by bound thermal neutrons (tens to hundreds of km), knowledge of the composition-dependent thermal-neutron fluxes, as well as the statistical nature of their location of origin needs to be known.

Future work to be accomplished should be focused on simulations to estimate the effects on $\tau_n$ of current uncertainties in atmospheric composition. These simulations will be done using currently known atmospheric compositions for Venus and Earth [e.g., 13]. If the expected systematic uncertainties are sufficiently large, then it will be determined what kind of mitigations exist, if any, for reducing the corresponding systematic uncertainties. For the landed lunar case, future work includes using current knowledge of measured thermal neutron fluxes, and carrying out simulations for different landing sites to quantify the expected systematic uncertainties.

*Atmospheric/surface temperature:* Since thermal neutrons are in thermal equilibrium with their surrounding material (either solid surface or gaseous atmosphere), their measured flux will depend on the temperature of that material. On the Moon, surface temperatures can range from 100 K at night to 400 K in the daytime [28]. However, temperatures at tens of cm below the lunar surface, where thermal neutrons are dominantly produced, have a smaller variation [29] than surface temperatures. Atmospheric temperatures on Venus for the altitudes where thermal neutrons are dominantly produced (60 – 75 km; see Figure S4 of [11]) can range from 200 K to ~350 K [13, 30]. The temperature of Earth's atmosphere is well known (e.g., [13]).

Based on particle transport simulations [12, 31], thermal-neutron count rates can vary by 4% to 6% from 100 K to 400 K. Since this variation is comparable to or larger than the variations for neutron lifetime, temperature variations either need to be limited and/or included in the modeling. Ways to limit temperature variations for orbital missions include using orbit parameters where temperature variations can be minimized, such as a sun synchronous orbit (e.g., always orbit over the dawn/dusk terminator). However, because the measured thermal neutrons will originate from a wide range of longitudes, this effect cannot be completely negated via orbit selection. Data can also be segregated into different temperatures when the data were gathered, and then corrections can be applied for these different temperatures. Such corrections can be carried out using first-principle corrections with known temperature variations, as well as empirical, data-driven corrections. In general, future work will use known temperature information, and after limitations are applied for a given mission scenario, investigate what level of uncertainties are caused by temperature variations, and the degree to which these variations can be corrected and/or included in the simulated count rates.

*Background signals:* As in any experiment, background signals need to be minimized where possible, and corrected for when present. For space-based neutron detectors, backgrounds can be present as both non-neutron and neutron signatures. Non-neutron signatures are most likely due to charged particles and resulting photon radiation that deposit energy in the neutron sensor. Li glass scintillators manifest a non-zero continuum background due to GCRs that needs subtraction from the measured neutron signature [18]. In contrast, $^3$He sensors are relatively insensitive to charged particles and energetic photons (x- and gamma-rays), and thus typically exhibit a very low non-neutron background [15, 26]. In addition to the nominal GCRs that are present at all times, there are also bursts of energetic particles that can cause large temporary backgrounds. These bursts come from solar energetic particle (SEP) events, and particle bursts from magnetospheric events within a planetary magnetosphere. For the locations considered here, all are susceptible to SEPs. However, Earth orbiting experiments might also need to contend with particle bursts depending on where in Earth's magnetosphere it is flown. In all these cases, particle bursts are usually dealt with by not using data acquired during the burst. Historically, removal of data contaminated by SEP and other particle events lead to a loss of approximately 15% calendar time of a given mission [24, 25].

A second source of background are direct neutron backgrounds. GCRs produce high-energy fast neutrons ($E_n > 0.5$ MeV) via interactions with spacecraft and sensor materials, and these fast neutrons are routinely measured in space-based neutron experiments [32, 33]. However, the amount of mass in most spacecraft is sufficiently low such that fast neutrons are not efficiently scattered to thermal energies [18, 26]. Thus, the background thermal-neutron count rate is quite low. Even so, for a dedicated space-based neutron lifetime experiment, effort should be made to minimize the surrounding mass around the sensor, which will minimize the overall neutron background counts. This argues for a dedicated small-satellite mission, as opposed to a hosting a neutron lifetime instrument on a larger spacecraft. In terms of background corrections, these can be made to the data by peak fitting individually measured spectra. Orbital design can also be used such that an elliptical orbit with sufficiently high-altitude apoapsis allows for a regular cadence of background measurements that can then be subtracted from the primary planetary measurements. The effect of such an elliptical orbit on the overall statistics is a trade that would need to be investigated. On one hand, a lower altitude periapsis would result in an increased count rate that could offset the loss in statistics from the lower count rate at higher altitudes. Alternatively, such an orbit could result in an overall loss of statistics where a "duty cycle" effect would need to be taken into account for the overall mission lifetime.

*Sensor efficiency response:* Knowledge of the angle and energy dependent sensor efficiency is needed to convert the measured count rate to a derived neutron flux, from which the neutron lifetime is determined. Uncertainties in the knowledge of this sensor response can potentially be a driving factor leading to unacceptably large uncertainties. Typically, a full sensor response is determined using a combination of particle transport modeling using codes like MCNPX or GEANT, and benchmarking to pre-launch calibration data. To accurately understand the response of these sensors to neutrons in space, all effective spacecraft material needs to be accurately modeled, and not just the sensor material. This modeling can be challenging for full sized spacecraft that have meter-sized dimensions or greater. In addition, for any given scenario, the exact spacecraft composition and mass distribution may not be fully known. Nevertheless, full simulations of MESSENGER NS data showed that relative uncertainties of ~<0.5% were achievable when neutron arrival angles were restricted to directions that did not travel through the full spacecraft material [7, 18, 20, 21, 34]. However, a dedicated space-based neutron lifetime experiment would strive to have a smaller spacecraft that is more easily modeled, thus achieving more uniform accuracies for all orientations. For landed experiments, missions that minimize non-sensor mass will enable more accurate modeling.

*Spacecraft location, velocity, and orientation:* To accomplish space-based neutron lifetime measurements, spacecraft orbital parameters, or landed location, will need to be known with sufficient precision such that uncertainties in these values do not significantly affect the overall lifetime-measurement uncertainty. For orbital spacecraft, standard navigation techniques enable the reconstruction of orbital spacecraft location, velocity, and orientations to very high degrees of accuracy. As an example, the Lunar Reconnaissance Orbiter spacecraft, has a reported root square sum position accuracy of <50 m, and a radial position accuracy of <1 m [35]. For a thermal neutron with a speed of 2.2 km/s, this can translate to a timing uncertainty of order 5 to 20 ms, which is substantially lower than a targeted lifetime uncertainty of 1 s. Velocity uncertainties typically have even lower values of <1 cm/s. While simulations for specific mission scenarios can be carried out to investigate the effect of navigation uncertainties on neutron lifetime measurements, it is expected that the overall uncertainties will be very small. Along with position and velocity knowledge, orientation/pointing uncertainties also tend to be very small for typical spacecraft, with uncertainty values generally less than 1°. As with the other spacecraft navigation parameters, these uncertainties need to be factored into neutron lifetime simulations to determine their effect on the neutron lifetime measurements. Finally, for landed measurements on the Moon, landing location uncertainties are very small (<1 m). Even so, it is not clear how any lunar landing location uncertainties will translate into lifetime uncertainties, but it is expected they would have little impact to any overall neutron lifetime measurement.

## 5. Discussion and Summary

Here, we provide a discussion and summary of the results given in Sections 3 and 4, and describe paths forward for planning and possibly accomplishing a space-based neutron lifetime measurement. Based on the statistical uncertainty results of Figure 7, Venus is clearly the best location for making neutron lifetime measurements. For the same observation time, Venus provides an almost order-of-magnitude better statistical uncertainty than an Earth orbiting or Moon-landed experiment. Venus has other advantages such that its space environment is relatively benign since it has no significant magnetic field to generate energetic particles or location-dependent changes to the local GCR flux. In addition, compared to the Earth and Moon, its composition is relatively simple, being mostly a two-component atmosphere (e.g., $CO_2$ and $N_2$). However, work needs to be done to understand the extent to which uncertainties in Venus' atmospheric composition and temperature would propagate to uncertainties in a derived neutron lifetime. One of the major challenges for making neutron lifetime measurements at Venus is the difficulty in getting to Venus orbit. While multiple spacecraft have orbited Venus, it is not inexpensive, and the cost required for a dedicated mission could be prohibitive. However, recent NASA planetary missions have emphasized rideshare opportunities where smaller satellites are placed on the same launch vehicle as a larger mission. If a neutron lifetime spacecraft could be hosted on a mission already traveling to Venus (e.g., [36]), then this could reduce launch costs significantly, possibly enabling such a mission to be accomplished.

Earth orbiting measurements have poorer statistical performance than measurements at Venus, but still can accomplish statistically robust measurements in a few months, and achieve <1-s statistical precision in less than a year of operation. Measurements at Earth have the clear advantage that it is much easier to reach Earth orbit than Venus. In addition, to increase statistical precision, more detector area could be used. We note that detector area (or equivalently, total counts) scales as observation time. Thus, increasing the total detector area by a factor of two would decrease the needed observation time by the same factor of two for an equivalent statistical precision.

While there are advantages to Earth orbiting measurements, such measurements do present challenges. First, while Earth's atmospheric composition is well known, its variability (especially $H_2O$ content) could propagate to uncertainties in derived neutron lifetime, and such variabilities need to be assessed. Earth's magnetic field can also cause uncertainties due to energetic particle backgrounds (depending on orbit), as well as location dependent GCR fluxes. All these variabilities need to be understood as does how they can either be corrected and/or affect any final neutron lifetime uncertainties.

Orbital measurements at the Moon provide the poorest statistical uncertainties, and therefore provide the least favorable location for neutron lifetime measurements. However, landed lunar measurements are statistically equivalent to Earth orbiting measurements. While placing an instrument designed to measure the neutron lifetime on the lunar surface is inherently more difficult than getting into Earth orbit, the anticipated cadence of future missions to the Moon presents opportunities for a landed neutron lifetime experiment. In addition, the Moon's relatively benign space environment at a stable platform on its surface may enable measurements with lower systematic uncertainty than at the Earth (or maybe Venus). Thus, lunar landed measurements should be studied in more detail to assess their feasibility.

The next steps needed to assess the feasibility of a space-based neutron lifetime experiment are to work through all the expected systematic uncertainties given in Table 1 to determine the level to which they might unacceptably compromise a neutron lifetime measurement after all mitigations and/or corrections are carried out. In regards to the uncertainty categories in Table 1, we expect that the uncertainties in atmospheric composition and temperature are most likely to hinder the ability to make a competitive neutron lifetime measurement. Reasons for this assessment include a current lack of knowledge for how compositional and temperature uncertainties translate into an overall neutron lifetime uncertainty. Future simulation work is needed to gain this knowledge. If such compositional/temperature uncertainties do present a significant problem, we expect mitigations could include a mix of additional measurements of compositional/temperature parameters, as well as experimental and observational strategies to minimize the effect of such uncertainties. In terms of observational strategies, if certain temperature uncertainties are present for known observational parameters (e.g., planetary local time), then steps could be taken to only acquire data during times when such observational parameters are more favorable. Alternatively, subsets of data could be segregated for different observational parameters as part of an overall mission design strategy, where such a strategy would also take into account additional time needed to gather sufficient statistics at varied observational parameters. Another category of systematic uncertainty that could be a challenge is knowledge of the sensor efficiency response. While existing mission data shows a good ability to simulate measured datasets (e.g., to 0.5% relative uncertainty), more work needs to be done to better understand if this level of uncertainty in relative response is sufficient to make a competitive neutron lifetime measurement. If not, then work needs to be done to determine what level of sensor efficiency knowledge is needed, and further if such knowledge can be achieved with standard spaceflight instrumentation. Finally, future work can be done to identify other sources of systematic uncertainties not discussed here. If it is determined that these uncertainties can be brought to an acceptable level, then more detailed mission design studies can be carried out where specific requirements are identified, and the resources needed to accomplish a given mission scenario are determined.

In summary, we have shown that the statistical precision of a space-based neutron lifetime experiment can be competitive with existing laboratory-based measurements. The best near-Earth location to carry out such an experiment is in Venus orbit, but measurements can also be accomplished in Earth orbit or on the surface of the Moon.

**Acknowledgements:** This work was supported by the Independent Research & Development Program at Johns Hopkins University Applied Physics Laboratory. The authors thank the anonymous reviewer for insightful comments leading to a better paper.

**Tables**

**Table 1.** Categories of systematic uncertainties and mitigation strategies for orbital and landed measurements.

| Uncertainties | Mitigation for Orbital Measurements | Mitigation for Landed Measurements |
|---|---|---|
| Time variable cosmic-ray flux | Separate radiation sensor; self-normalization with slow spinning spacecraft | Separate radiation sensor; self-normalization with Cd-covered and non-Cd-covered sensor |
| Atmospheric/surface composition | Knowledge of elemental composition; thick atm. provides less spatial variability | Single location reduces variability; account for long travel distance of thermal neutrons |
| Atmospheric/surface temperature | Knowledge of surface temp.; multiple measurements of similar temps.; orbits to limit temp. variability | Knowledge of surface temp.; multiple measurements of similar temps. |
| Background signals | Avoid radiation belts/planetary mag. fields; small spacecraft has decreased background | Reduce non-sensor mass. |
| Sensor efficiency response | Relative accuracy of surface-to-spacecraft model achieves <0.5%; small spacecraft easier to model; slow spinning spacecraft cross check on sensor-to-sensor response | Small instrument enables more accurate modeling |
| Spacecraft location, velocity, and orientation | Known with high precision based on standard space navigation techniques | Landed accuracy will be known with high precision |

**Figures**

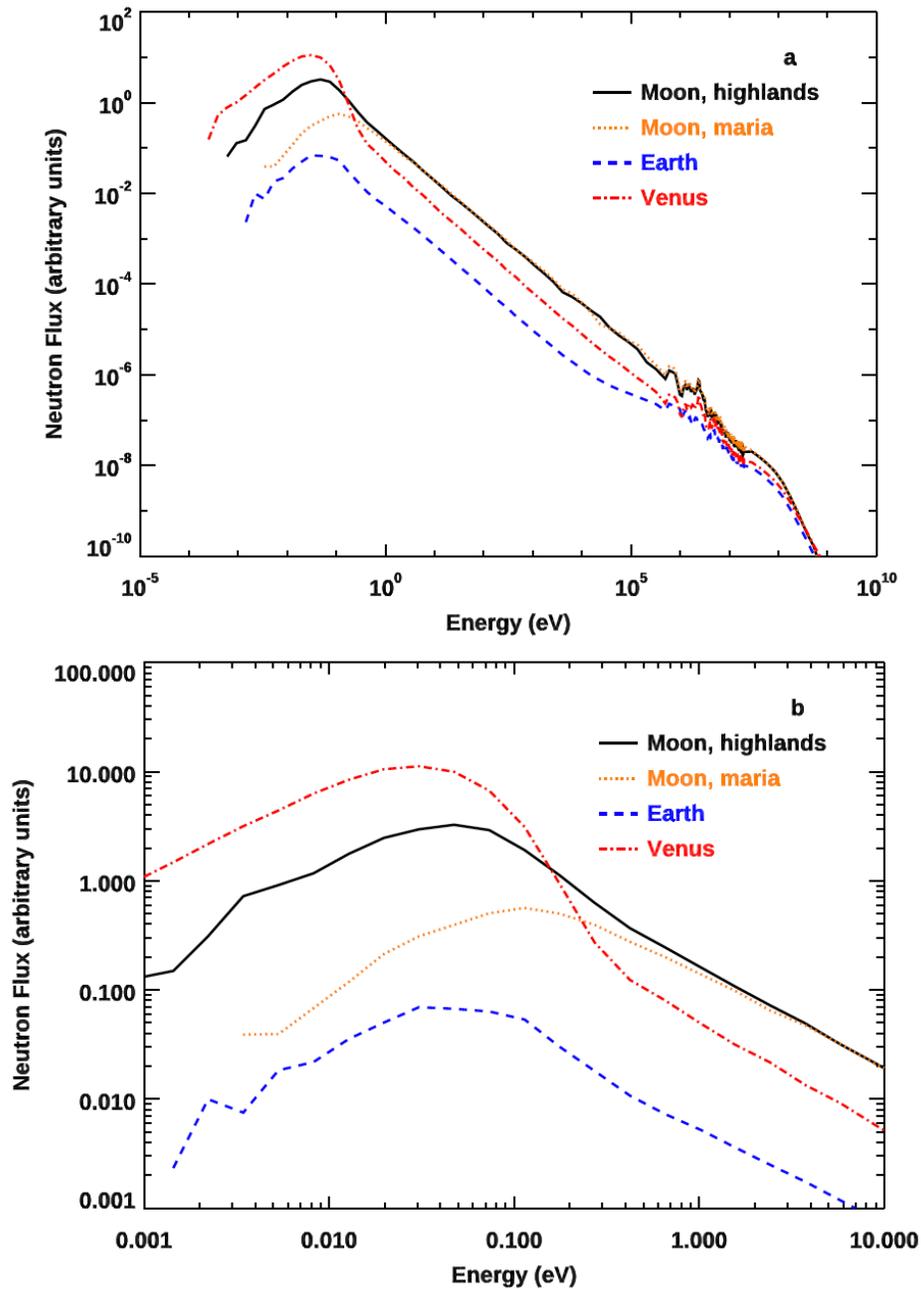

**Figure 1.** Simulated neutron fluxes for four different planetary surfaces for all energies from thermal to fast (a), and only for thermal and low-energy epithermal energies (b). Moon highlands and maria are shown with black/solid and orange/dotted, respectively; Venus shown in red/dot-dashed; Earth shown in blue/dashed.

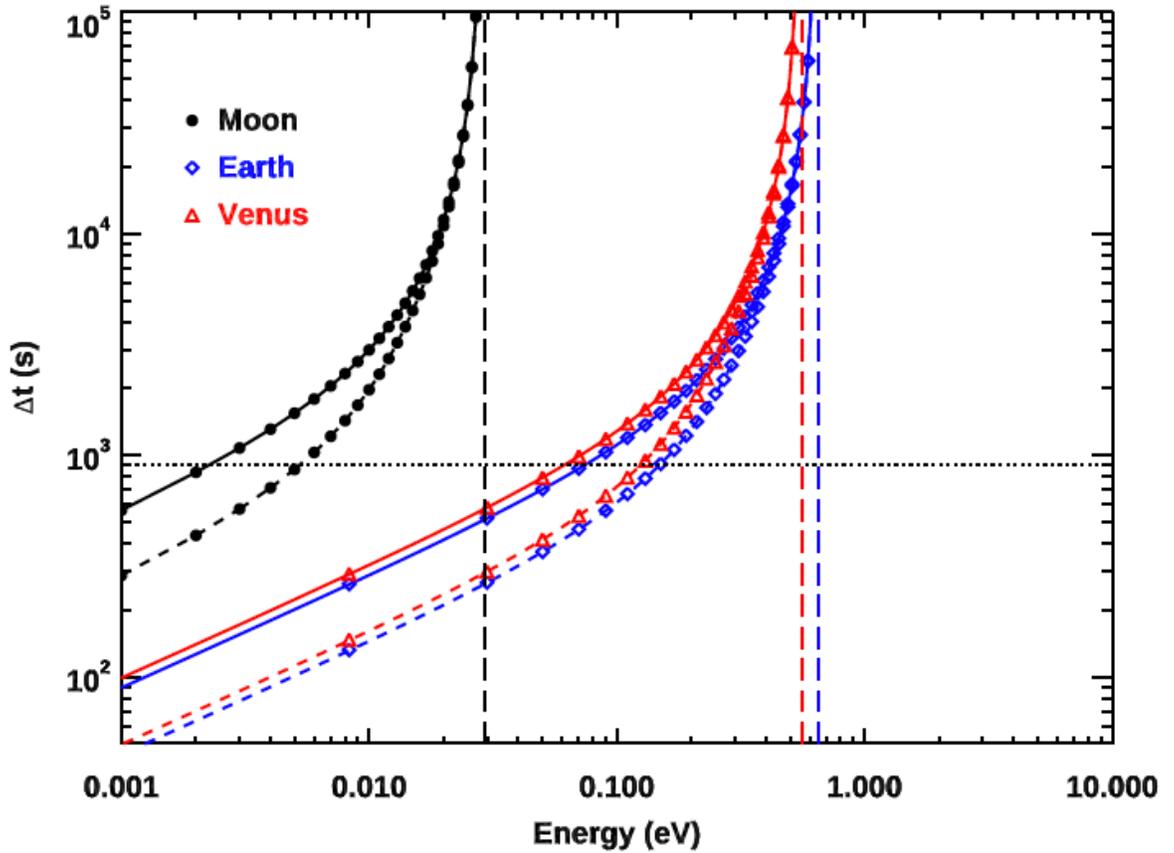

**Figure 2.** Surface-to-surface neutron return time, Δ*t*, versus neutron energy for the Moon (black circles), Earth (blue diamonds) and Venus (red triangles). Solid lines show Δ*t* for 0° launch angle (θ); short dashed lines show Δ*t* for 60° launch angle. Vertical dashed lines show the gravitational bounding potential for each planet. Horizontal dotted line shows the currently accepted neutron lifetime of 880 s.

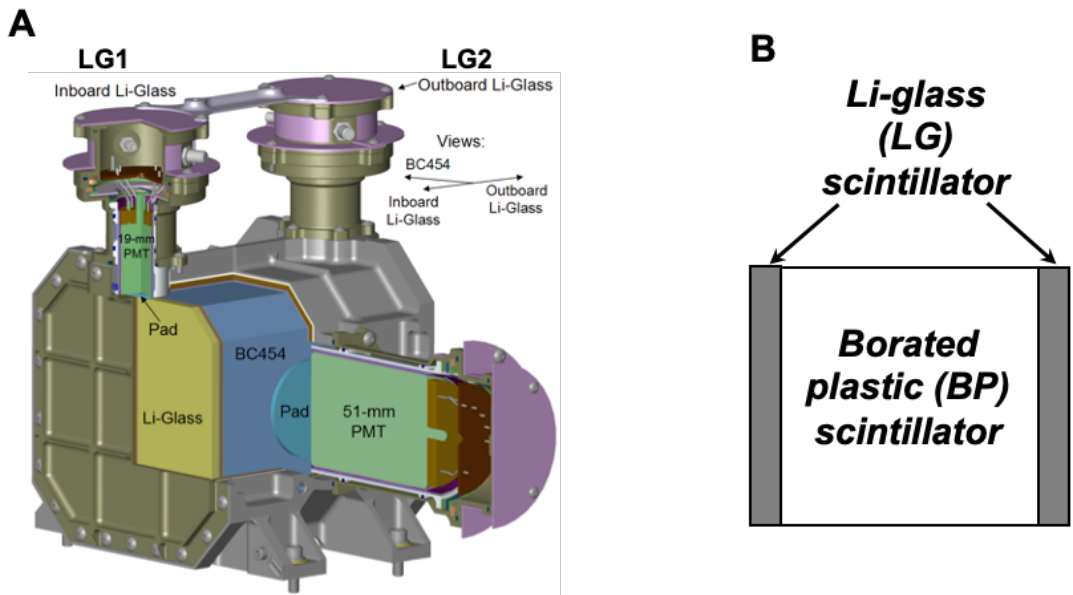

**Figure 3.** (A) Engineering drawing of the MESSENGER Neutron Spectrometer (NS); and (B) a simplified schematic of the placement of the three NS sensor components. BC454 in (A) refers to the designation of the borated plastic scintillator. The inboard Li-glass sensor is labeled LG1, and its normal vector points in the spacecraft +x direction; the outboard Li-glass sensor is labeled LG2 and its normal vector points in the spacecraft -x direction.

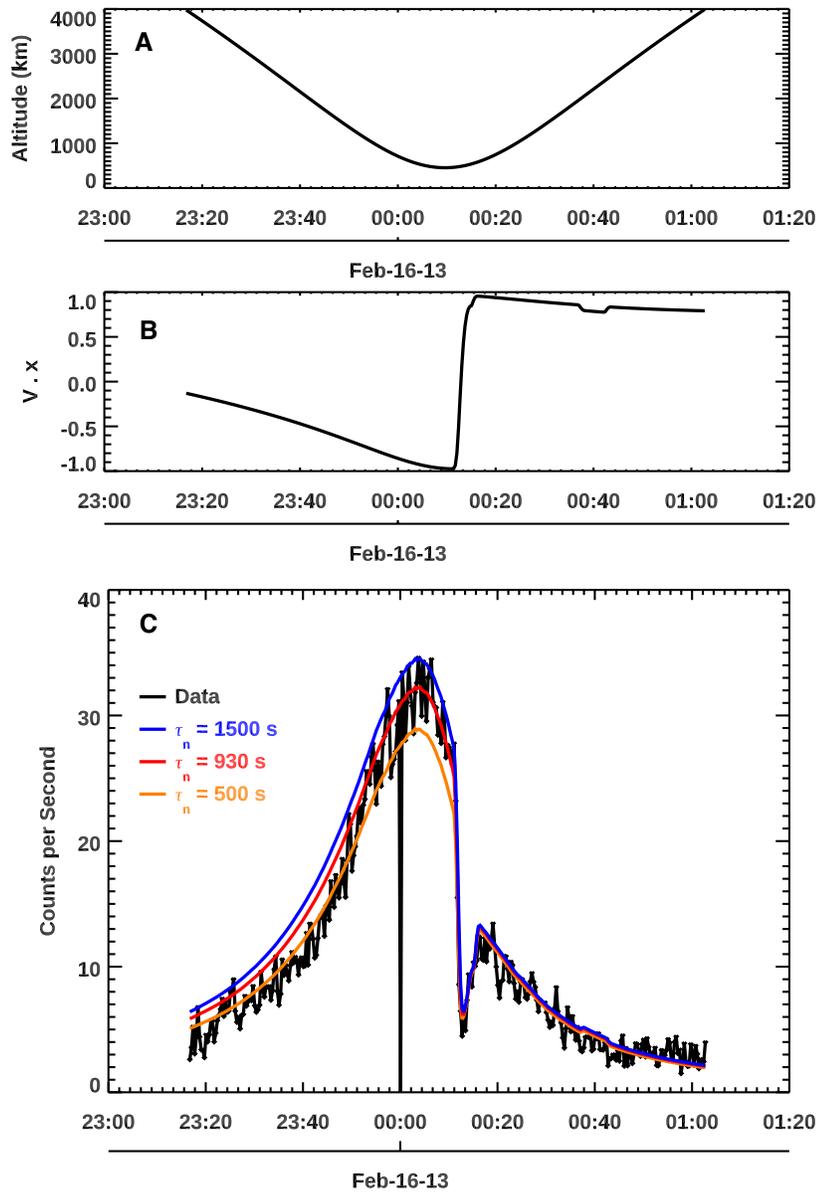

**Figure 4.** (A) MESSENGER altitude, (B) spacecraft velocity-direction dot product with the spacecraft x-axis direction, and (C) NS measured and modeled count rates from the LG2 sensor for one orbit around Mercury on February 16, 2013. The orbit periapsis occurs for an altitude of 450 km at a time of 00:09:24 UTC. Prior to the periapsis, the spacecraft was oriented to produce a thermal-neutron enhancement (maximum Doppler effect, $\boldsymbol{V} \cdot x \cong -1$) for the LG2 sensor. At the time around periapsis, the spacecraft executed a rotation such that the LG2 sensor direction

was changed to give a thermal-neutron decrease (minimum Doppler effect, $\boldsymbol{V} \cdot \boldsymbol{x} \cong +1$). The measured LG2 count rate shows a maximum around the spacecraft periapsis. The large count-rate decrease after periapsis is due to both the larger spacecraft altitude and the thermal-neutron decrease from the minimum Doppler effect. Modeled count rates were calculated for three different neutron lifetimes of 500, 930, and 1500 s, and were normalized to the count rates at periapsis. Modeled count rates show a large separation for different values of $\tau_n$ with maximum Doppler effect, and a small separation with the minimum Doppler effect.

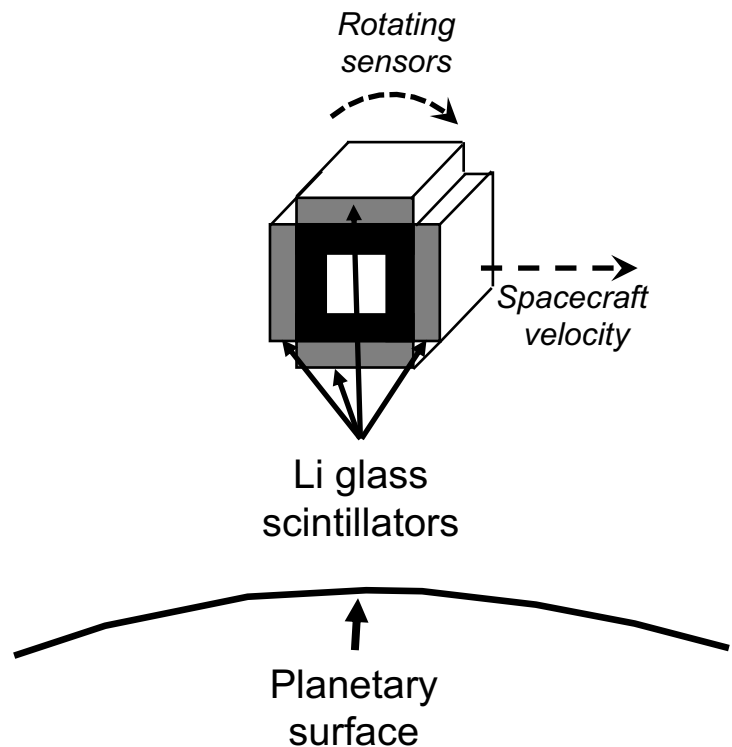

**Figure 5.** Notional sensor/spacecraft arrangement for an orbital neutron lifetime experiment. The experiment uses four Li-glass scintillators with a size of 10 cm by 10 cm by 4 mm. The sensor package is spin stabilized with a rotation axis that goes through the center of the four sensors. The sensors are arranged at right angles to each other with the velocity such that each sensor cycles from maximum to minimum Doppler effect.

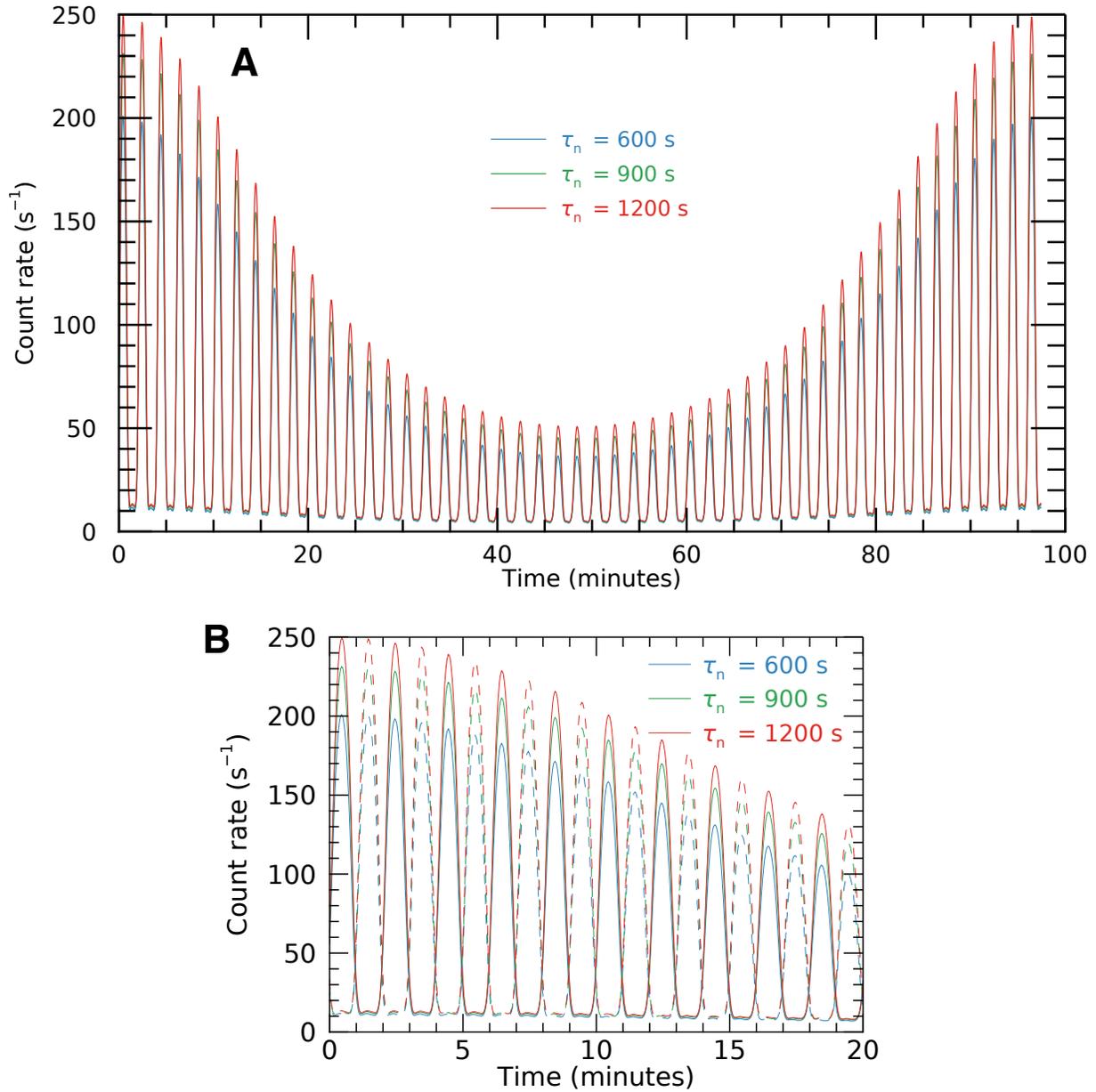

**Figure 6.** Simulated neutron count rate for a 250 km by 750 km elliptical orbit around Venus. (A) shows one orbit for a single sensor in the notional package shown in Figure 5. Different colors illustrate the count rate for different modeled $\tau_n$ values. (B) shows 20 minutes of an orbit where opposing sensors are shown in solid and dashed lines. In both plots, the highest count rates are from $\tau_n$ = 1200 s (red); the mid-range counts are from $\tau_n$ = 900 s (green); the lowest count rates are from $\tau_n$ = 600 s (blue).

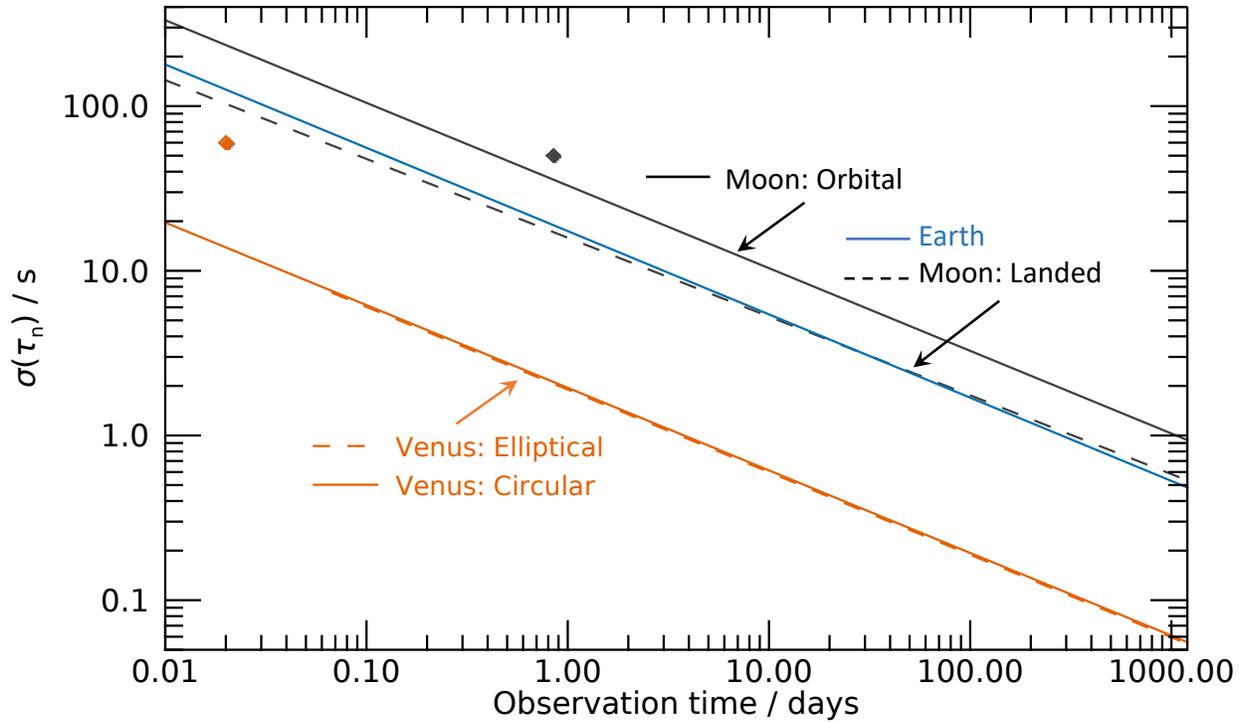

**Figure 7.** Statistical precision, $\sigma(\tau_n)$, of a neutron lifetime measurement versus observation time. Orbital measurements from circular orbits about the Moon, Earth, and Venus are shown in solid black, blue, and orange respectively. Landed measurements on the Moon are shown in black dashed. Orbital measurements from a 250 km by 750 km elliptical orbit about Venus are shown in orange dashed. The orange diamond shows the $\tau_n$ measurement at Venus from [4]; the black diamond shows the $\tau_n$ measurement at the Moon from [5].

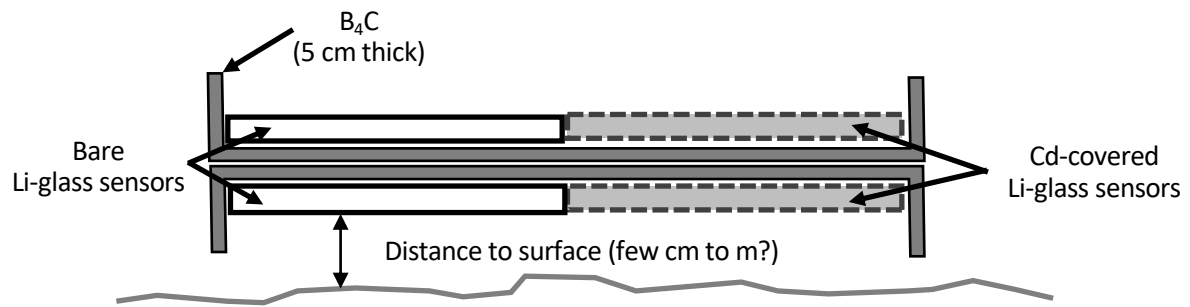

**Figure 8.** Notional sensor arrangement for a lunar landed neutron lifetime experiment. The experiment uses four Li-glass sensors with a size of 10 cm by 10 cm by 4 mm. One pair of sensors look upwards, and one pair look downwards. Each pair is shielded by a 5-cm thick layer of $^{10}$B-enriched $B_4C$. For each pair, one sensor is covered in a layer of Cd and one sensor is bare. Each Cd-covered sensor provides a measure of epithermal neutrons and each bare sensor measures thermal plus epithermal neutrons. The count rate difference between each sensor within the pair provides a measure of thermal neutrons. The distance to the surface can be anywhere from a few cm to approximately a meter.

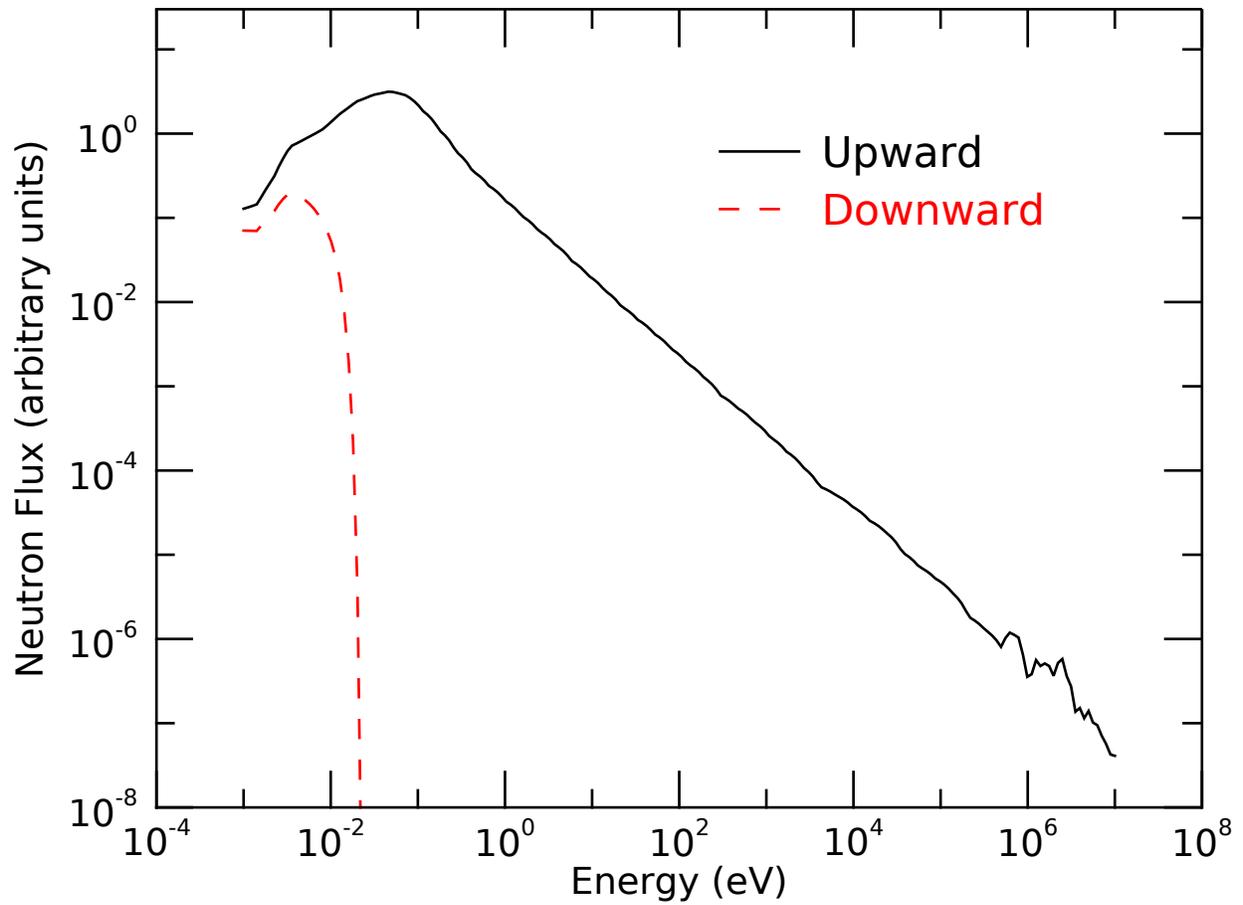

**Figure 9.** Upward (black solid) and downward (red dashed) neutron flux as seen from the lunar surface for a highlands-type lunar material.

# References

1. Wietfeldt, F.E., *Measurements of the Neutron Lifetime.* Atoms, 2018. **6**(4).
2. Feldman, W.C., G.F. Auchampaugh, and D.M. Drake, *A Technique To Measure The Neutron Lifetime From Low-Earth Orbit.* Nuclear Instruments & Methods in Physics Research Section a-Accelerators Spectrometers Detectors and Associated Equipment, 1990. **287**(3): p. 595-605.
3. Feldman, W., et al., *Chemical information content of lunar thermal and epithermal neutrons.* Journal of Geophysical Research-Planets, 2000. **105**(E8): p. 20347-20363.
4. Wilson, J.T., et al., *Space-based measurement of the neutron lifetime using data from the neutron spectrometer on NASA's MESSENGER mission.* Physical Review Research, 2020. **2**: p. 023316.
5. Wilson, J.T., et al., *Measurement of the Neutron Lifetime Using Data from the Neutron Spectrometer on NASA's Lunar Prospector Mission.* Physical Review C: Nuclear Physics, 2020: p. in review.
6. Maurice, S., et al., *Mars Odyssey neutron data: 1. Data processing and models of water-equivalent-hydrogen distribution.* Journal of Geophysical Research, 2011. **116**(E11): p. E11008-E11008.
7. Lawrence, D.J., et al., *Evidence for water ice near Mercury's north pole from MESSENGER Neutron Spectrometer measurements.* Science, 2013. **339**(6117): p. 292-296.
8. Feldman, W.C., et al., *Fluxes of fast and epithermal neutrons from Lunar Prospector: Evidence for water ice at the lunar poles.* Science, 1998. **281**(5382): p. 1496-1500.
9. Feldman, W., et al., *Global distribution of neutrons from Mars: Results from Mars Odyssey.* Science, 2002. **297**(5578): p. 75-78.
10. Prettyman, T.H., et al., *Elemental Mapping by Dawn Reveals Exogenic H in Vesta's Regolith.* Science, 2012. **338**(6104): p. 242-246.
11. Peplowski, P.N., D.J. Lawrence, and J.T. Wilson, *Chemically distinct regions of Venus's atmosphere revealed by measured N-2 concentrations.* Nature Astronomy, 2020.

12. Lawrence, D.J., et al., *Improved modeling of Lunar Prospector neutron spectrometer data: Implications for hydrogen deposits at the lunar poles.* Journal of Geophysical Research, 2006. **111**(E8): p. E08001.
13. Lodders, K. and B.J. Fegley, *The Planetary Scientist's Companion*. 1998, New York: Oxford University Press. 371.
14. Feldman, W.C., et al., *Gravitational effects on planetary neutron flux spectra.* Journal of Geophysical Research, 1989. **94**(B1): p. 513-513.
15. Feldman, W., et al., *Gamma-Ray, Neutron, and Alpha-Particle Spectrometers for the Lunar Prospector mission.* Journal of Geophysical Research: Planets, 2004. **109**(E7).
16. Goldsten, J.O., et al., *The MESSENGER Gamma-Ray and Neutron Spectrometer.* Space Science Reviews, 2007. **131**(1-4): p. 339-391.
17. Feldman, W.C. and D.M. Drake, *A Doppler Filter Technique To Measure The Hydrogen Content Of Planetary Surfaces.* Nuclear Instruments & Methods in Physics Research Section a-Accelerators Spectrometers Detectors and Associated Equipment, 1986. **245**(1): p. 182-190.
18. Lawrence, D.J., et al., *Identification and measurement of neutron-absorbing elements on Mercury's surface.* Icarus, 2010. **209**(1): p. 195-209.
19. Solomon, S.C. and B.J. Anderson, *The MESSENGER Mission: Science and Implementation Overview*, in *Mercury: The View after MESSENGER*, L.R.N. Sean C. Solomon, Brian J. Anderson, Editor. 2018, Cambridge University Press: Cambridge, United Kingdom. p. 1 - 29.
20. Peplowski, P.N., et al., *Geochemical terranes of Mercury's northern hemisphere as revealed by MESSENGER neutron measurements.* Icarus, 2015. **253**: p. 346-363.
21. Wilson, J.T., et al., *MESSENGER Gamma Ray Spectrometer and Epithermal Neutron Hydrogen Data Reveal Compositional Differences Between Mercury's Hot and Cold Poles.* Journal of Geophysical Research-Planets, 2019. **124**(3): p. 721-733.
22. Lawrence, D.J., et al., *Measuring the Elemental Composition of Phobos: The Mars-moon Exploration with GAmma rays and NEutrons (MEGANE) Investigation for the Martian Moons eXploration (MMX) Mission.* Earth and Space Science, 2019. **6**(12): p. 2605-2623.

23. Lawrence, D.J., et al., *The Psyche Gamma-Ray and Neutron Spectrometer: Update on Instrument Design and Measurement Capabilities*, in *50th Lunar and Planetary Science Conference*. 2019: Houston, Texas. p. Abstract #1544.
24. Lawrence, D.J., S. Maurice, and W.C. Feldman, *Gamma-ray measurements from Lunar Prospector: Time series data reduction for the Gamma-Ray Spectrometer.* Journal of Geophysical Research, 2004. **109**(E7): p. E07S05-E07S05.
25. Lawrence, D.J., et al., *Galactic cosmic ray variations in the inner heliosphere from solar distances less than 0.5AU: Measurements from the MESSENGER Neutron Spectrometer.* Journal of Geophysical Research: Space Physics, 2016. **121**.
26. Maurice, S., et al., *Reduction of neutron data from Lunar Prospector.* Journal of Geophysical Research, 2004. **109**(E7): p. E07S04.
27. Prettyman, T.H., et al., *Elemental composition of the lunar surface: Analysis of gamma ray spectroscopy data from Lunar Prospector.* Journal of Geophysical Research, 2006. **111**(E12): p. E12007-E12007.
28. Williams, J.P., et al., *The global surface temperatures of the moon as measured by the diviner lunar radiometer experiment.* Icarus, 2017. **283**: p. 300-325.
29. Hayne, P.O., et al., *Global Regolith Thermophysical Properties of the Moon From the Diviner Lunar Radiometer Experiment.* Journal of Geophysical Research-Planets, 2017. **122**(12): p. 2371-2400.
30. Limaye, S.S., et al., *Venus Atmospheric Thermal Structure and Radiative Balance.* Space Science Reviews, 2018. **214**(5).
31. Little, R., et al., *Latitude variation of the subsurface lunar temperature: Lunar Prospector thermal neutrons.* Journal of Geophysical Research-Planets, 2003. **108**(E5).
32. Feldman, W.C., et al., *Fast neutron flux spectrum aboard Mars Odyssey during cruise.* Journal of Geophysical Research-Space Physics, 2002. **107**(A6).
33. Rodgers, D.J., et al., *Neutrons and energetic charged particles in the inner heliosphere: Measurements of the MESSENGER Neutron Spectrometer from 0.3 to 0.85 AU.* Journal of Geophysical Research: Space Physics, 2015. **120**(2): p. 841-854.
34. Lawrence, D.J., et al., *Compositional terranes on Mercury: Information from fast neutrons.* Icarus, 2017. **281**: p. 32-45.

35. Beckman, M. *Mission Design for the Lunar Reconnaissance Orbiter*. in *29th Annual AAS Guidance And Control Conference*. 2006. Breckenridge, Colorado.
36. Smrekar, S.E., et al., *VERITAS (Venus Emissivity, Radio Science, Insar, Topography And Spectroscopy): A Proposed Discovery Mission*, in *51st Lunar and Planetary Science Conference*. 2020. p. Abstract #1499.